\documentclass[aps, reprint, prl, superscriptaddress, 10pt]{revtex4-1}
\usepackage{graphicx}
\usepackage{epstopdf}
\usepackage[breaklinks,colorlinks,citecolor=blue,linkcolor=red,urlcolor=blue]{hyperref}
\usepackage{amsmath}
\usepackage{setspace}

\begin{document}

\title{Magnetic excitations in the \(S=\frac{1}{2}\) antiferromagnetic-ferromagnetic chain compound BaCu$_2$V$_2$O$_8$ at zero and finite temperature}

\author{E. S. Klyushina}
\affiliation{Helmholtz-Zentrum Berlin f{\"u}r Materialien und Energie, 14109 Berlin, Germany} 
\affiliation{Institut f{\"u}r Festk{\"o}rperphysik, Technische Universit{\"a}t Berlin, 10623 Berlin, Germany} 
\author{A. C. Tiegel}
\affiliation{Institut f{\"u}r Theoretische Physik, Georg-August-Universit{\"a}t G{\"o}ttingen, 37077 G{\"o}ttingen, Germany}
\author{B. Fauseweh}
\affiliation{Lehrstuhl f{\"u}r Theoretische Physik I, Technische Universit{\"a}t Dortmund, 44221 Dortmund, Germany}
\author{A.T.M.N. Islam}
\affiliation{Helmholtz-Zentrum Berlin f{\"u}r Materialien und Energie, 14109 Berlin, Germany} 
\author{J. T. Park}
\affiliation{Heinz Maier-Leibnitz Zentrum, TU M{\"u}nchen, 85747 Garching, Germany} 
\author{B. Klemke}
\affiliation{Helmholtz-Zentrum Berlin f{\"u}r Materialien und Energie, 14109 Berlin, Germany} 
\author{A. Honecker}
\affiliation{Laboratoire de Physique Th\'eorique et Mod\'elisation, CNRS UMR 8089, Universit\'e de Cergy-Pontoise, 95302 Cergy-Pontoise Cedex, France} 
\author{G. S. Uhrig}
\affiliation{Lehrstuhl f{\"u}r Theoretische Physik I, Technische Universit{\"a}t Dortmund, 44221 Dortmund, Germany}
\author{S. R. Manmana}
\affiliation{Institut f{\"u}r Theoretische Physik, Georg-August-Universit{\"a}t G{\"o}ttingen, 37077 G{\"o}ttingen, Germany} 
\author{B. Lake}
\affiliation{Helmholtz-Zentrum Berlin f{\"u}r Materialien und Energie, 14109 Berlin, Germany} 
\affiliation{Institut f{\"u}r Festk{\"o}rperphysik, Technische Universit{\"a}t Berlin, 10623 Berlin, Germany} 

\begin{abstract}
Unlike most quantum systems which rapidly become incoherent as temperature is raised, strong correlations persist at elevated temperatures in $S=1/2$ dimer magnets, as revealed by the unusual asymmetric lineshape of their excitations at finite temperatures. Here we quantitatively explore and parameterize the strongly correlated magnetic excitations at finite temperatures using the high resolution inelastic neutron scattering on the model compound BaCu$_2$V$_2$O$_8$ which we show to be an alternating antiferromagnetic-ferromagnetic spin$-1/2$ chain. Comparison to state of the art computational techniques shows excellent agreement over a wide temperature range. Our findings hence demonstrate the possibility to quantitatively predict coherent behavior at elevated temperatures in quantum magnets.
\end{abstract}

% insert suggested PACS numbers in braces on next line
%\pacs{}

\date{\today}  

\maketitle

\par In the study of unconventional states of matter, quantum magnetic materials with their strong correlations play a crucial role \cite{Balents:2010ve,book_frustratedspins,book_HFMTrieste,springer_quantummagnetism,
fazekas}. 
Quantum mechanical coherence and entanglement are intrinsic to these systems, both being relevant for potential applications in quantum devices \cite{Amico:2008en,nielsenchuang}. 
However, the question arises for their persistence when increasing temperature. Intuitively, one expects temperature to suppress quantum behavior, as typically encountered in the study of quantum criticality \cite{Sachdev}. 
Interestingly, this is not always the case, and in certain systems, e.g. in the presence of disorder, coherent behavior is not simply suppressed by temperature, but rather an interesting interplay develops \cite{NaturePhysicsAltshuler,Mohseni2008}, which can lead to counterintuitive behavior such as the increase of conductance through molecules with temperature \cite{PhysRevLett.109.056801}. 
\par Another example is the extraordinary coherence of the magnetic excitations at elevated temperatures. This was theoretically predicted for 1-dimensional (1D) gapped quantum dimer antiferromagnets (AFM) by using integrable quantum field theory \cite{a7a} and was experimentally confirmed on the strongly dimerized spin$-1/2$ AFM alternating chain compound copper nitrate, which has a spin-singlet ground state and gapped triplet excitations (henceforth referred to as triplons \cite{Schmidt2003}) confined within a narrow band  \cite{a6}. Here, the triplons interact strongly via the AFM interdimer coupling and also via an effective repulsive interaction due to the hard-core constraint. 
The resulting strong correlations lead to the experimentally observed asymmetric broadening of the lineshape with temperature \cite{a6,a6a}.  So far, such experimental data was compared to exact diagonalization data from small systems and to results from low-temperature expansion around the strongly dimerized limit of Heisenberg spin-$1/2$ chains \cite{a7,a8}. Further experimental studies revealed that the strongly correlated behavior at elevated temperatures is not restricted to 1D systems. It was recently observed that the lineshape in the 3-dimensional (3D) coupled-dimer antiferromagnet Sr$_3$Cr$_2$O$_8$ also becomes asymmetric and increasingly weighted towards the center of the band as temperature increases  \cite{a9,a10}. So far, no reliable theoretical approaches on the microscopic level are available which capture large systems beyond the limit of strong dimerization. The development of such techniques is crucial to provide a quantitative description of the strongly correlated behavior at finite temperatures.

\par
The scope of this Letter is to report the comparison of two currently developed theoretical approaches with quantitative predictive power to  experimental data. These approaches are based on matrix product states (MPS) or density-matrix renormalization group (DMRG) techniques \cite{White_1992,White_1993,Schollwoeck_2011} and on the diagrammatic Br\"uckner approach on top of Continuous Unitary Transformations (DBA-CUT) \cite{Fauseweh2014,Fauseweh2015}, respectively. They provide an accurate description for the strongly correlated behavior of the magnetic excitations at finite temperatures in the dimer compound BaCu$_2$V$_2$O$_8$. High resolution inelastic neutron scattering (INS) measurments are compared with the theoretical approaches.  The analysis of the experimental and theoretical results reveals accurate quantitative agreement between the experimentally observed and the theoretically predicted strongly correlated behavior at finite temperature. This is our first key result. Because the couplings in BaCu$_2$V$_2$O$_8$ have been strongly debated in the literature our second key result is to deduce the Hamiltonian of this compound and show that 
\begin{figure}[ht]
 \centering
 \includegraphics [width=\linewidth]{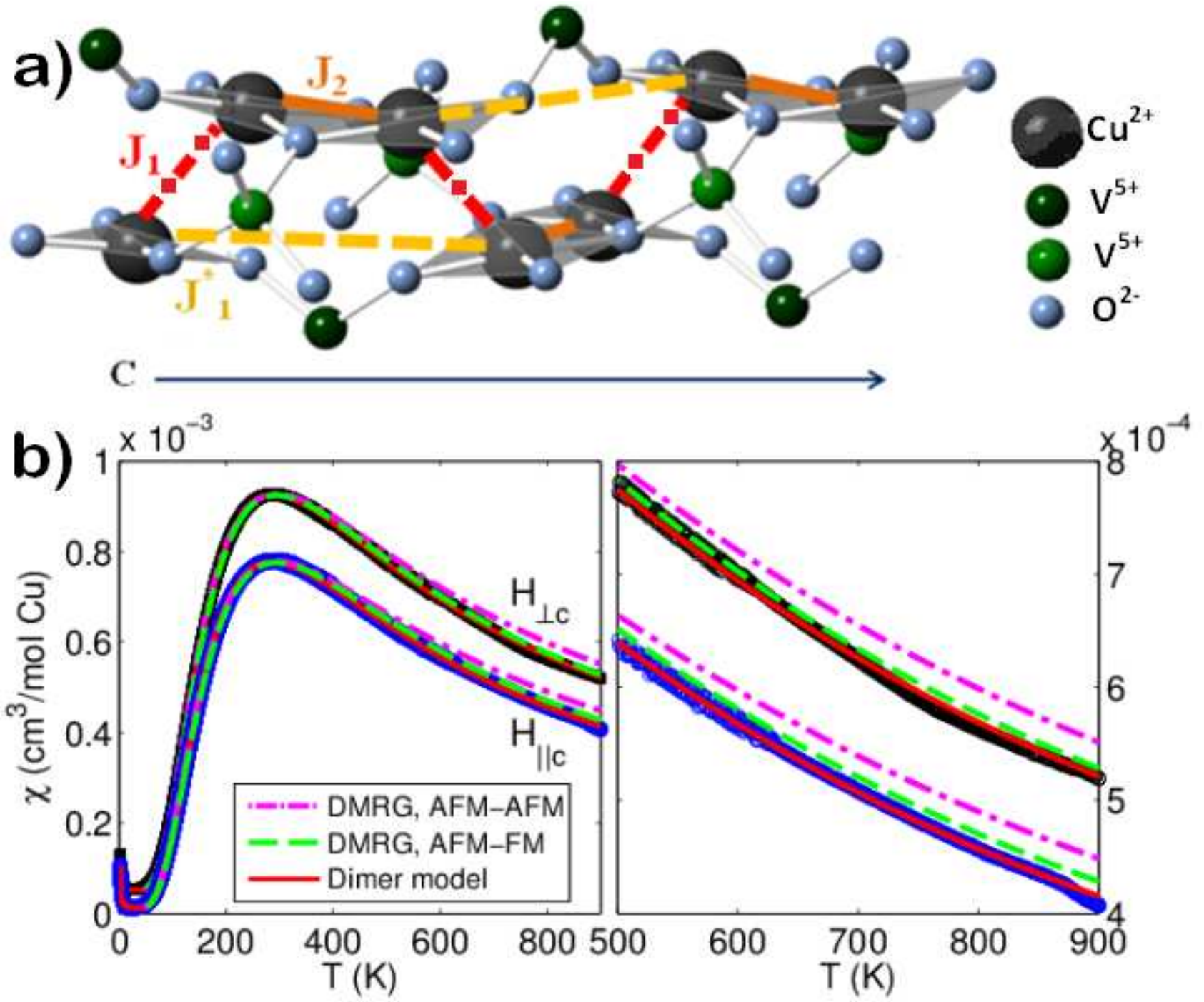}
 \caption{(Color online) (a) Crystal structure of BaCu$_2$V$_2$O$_8$ (the Ba$^{2+}$ are omitted) showing  the two proposed models of the dimerized chain arrangement along the $c$-axis: first model with exchange paths $J_1^*$ (dashed line) and $J_2$ (solid line) resulting in two independent non-interacting dimerized linear chains \cite{a12}; second model with exchange paths $J_1$ (dash-dotted line) and $J_2$ (solid line) leading to a single dimerized screw chain \cite{a16}. (b) $\chi _{DC}$ as a function of temperature for a 1 T magnetic field applied parallel and perpendicular to the $c$-axis. The solid line is the coupled dimer model \cite{a21,a22} where the $g$-factors and exchange constants were fitted ($J_{\mathrm{intra}} =39.8\pm 0.13$~meV, $J_{\mathrm{inter}}=-9.87\pm2.64$~meV, $g_{|| c} = 2.09$, $g_{\perp c} = 2.27$). The dashed and dash-dotted lines are DMRG results with the exchange constants fixed to the values for the AFM-AFM and AFM-FM models, respectively; only the $g$-factors were fitted. All fits included additional terms to account for paramagnetic impurities and van Vleck susceptibility. The AFM-FM model yields an anisotropic $g$-factor $g_{|| c}=2.14\pm0.015$ and $g_{\perp c} =2.29\pm0.015$.
In BaCu$_2$V$_2$O$_8$ the plaquettes contain the $c$-axis and rotate about it; $g_{|| \mathrm{plaquette}}=g_{|| c}=2.12 \pm 0.03$ and $g_{\perp \mathrm{plaquette}} =2g_{\perp c} - g_{|| c} =2.44 \pm 0.03$ in agreement with other cuprates with square planar coordination \cite{a19,a23,a24,a25,a26}.}
 \label{Fig1}
\end{figure} 
it is a highly dimerized antiferromagnetic-ferromagetic chain correcting the long-held view that the interdimer coupling is AFM or negligible. This observation implies our third key result that the presence of strongly correlated behavior in gapped dimer systems is independent on the sign of the interdimer
coupling.
\par {\it{Crystal structure.$-$}}BaCu$_2$V$_2$O$_8$ has a tetragonal crystal structure (space group I$\bar{4}$2d, lattice parameters $a=b=12.744$ \AA, $c=8.148$~\AA). The magnetic Cu$^{2+}$ ions ($S=1/2$) are coordinated by O$^{2-}$ ions in square-planar geometry and these CuO$_4$ plaquettes form edge-sharing pairs which rotate about the $c$-axis and are oriented with the $c$-axis lying within the plaquettes (Fig.\ \ref{Fig1}(a)).  Previous $\chi_{DC}$ \cite{a12,a13,a11,a14}, specific heat\,\cite{a12} and $^{51}$V nuclear magnetic resonance \cite{a14,a15} measurements revealed a non-magnetic ground state with excitations above a gap of size $\Delta\approx 31.0-40.5$\,meV. This  implies that the Cu$^{2+}$ ions are coupled into dimers by a dominant AFM intradimer interaction ($J_{\mathrm{intra}}$), resulting in a spin-singlet ground state and gapped triplon excitations. The interdimer interaction ($J_{\mathrm{inter}}$) was previously assumed to be AFM with strength between 0\,\% and 20\,\% of the intradimer coupling \cite{a11,a12,a13,a14,a15}. 

The exchange paths responsible for the $J_{\mathrm{intra}}$ and $J_{\mathrm{inter}}$ coupling are strongly debated in the literature \cite{a11,a12,a13,a16}. Two  models for BaCu$_2$V$_2$O$_8$ have been suggested (Fig.\ \ref{Fig1}(a)). The first, which assumes the paths $J_{2}$ and $J_{1}^{\ast}$, gives rise to almost straight independent non-interacting dimerized double chains parallel to the $c$-axis \cite{a12}. The second, which consists of $J_1$ and $J_2$, couples the Cu$^{2+}$ ions into a single alternating screw chain \cite{a16}. Both suggest that the AFM $J_{\mathrm{inter}}$ arises via the superexchange path $J_2$ (Cu-O-Cu) \cite{a16} between the two Cu$^{2+}$ ions within the edge-sharing plaquettes while the $J_{\mathrm{intra}}$ is realized  via AFM super-superexchange path $J_{1}$ or $J_{1}^{\ast}$ (Cu-O-V-O-Cu) \cite{a16}.  
The second model is favored by two band structure investigations which predict that  $J_1$ and $J_2$ are both AFM with ratio $J_2 / J_1$ of 0.16 \cite{a16} or 0.05 \cite{a11}, while $J_{1}^{\ast}$ is much weaker. 

{\it{Methods.$-$}} Single crystals of BaCu$_2$V$_2$O$_8$ were grown in the Crystal Laboratory at the Helmholtz Zentrum Berlin f{\"u}r Materialien und Energie (HZB), using the traveling-solvent-floating-zone method \cite{a17}. $\chi _{DC}$ was measured using a superconducting quantum interference device at the Laboratory for Magnetic Measurements, HZB, over the temperature range 2-900 K.  Single crystal INS measurements were performed on the thermal triple-axis spectrometer PUMA \cite{Puma}. The magnetic excitation spectrum was mapped out at $T=5$~K using double-focused pyrolytic graphite (PG(002)) monochromator and analyzer with fixed final wavevector $k_f =2.662$~\AA$^{-1}$ giving an energy resolution of 2 meV. The lineshape of the excitations was measured at the dispersion minima (6,0,1) and (8,0,0), for temperatures in the range of 3.5-200 K using a double-focused Cu(220) monochromator and PG(002) analyzer with fixed $k_f =1.97$~\AA$^{-1}$ to give a higher energy resolution of 0.74 meV. The excitation spectra of BaCu$_2$V$_2$O$_8$ were calculated in the frequency-domain using DMRG-based Chebyshev expansions \cite{Weisse2006} at zero \cite{Holzner2011, Braun2014, Wolf2015} and finite temperature \cite{Tiegel2014, Tiegel2015}  taking into account the positions of the Cu$^{2+}$ ions \cite{SM}.  At finite temperature, this approach is combined with linear prediction \cite{Ganahl2014,Wolf2014a}. The diagrammatic Br{\"u}ckner approach was used to compute the thermal fluctuations of the strongly interacting hardcore bosons on top of the effective model obtained by a Continuous Unitary Transformation (DBA-CUT) \cite{Fauseweh2014,Fauseweh2015,SM}. Both calculations  were performed for the $S=1/2$ alternating chain Heisenberg Hamiltonian  
\begin{equation}
 {H = \sum_i \, J_{\mathrm{intra}}\,\mathbf{S}_{i,1}\cdot\mathbf{S}_{i,2} + J_{\mathrm{inter}}\, \mathbf{S}_{i,2}\cdot\mathbf{S}_{i+1,1}. \label{eq: Hamiltonian}}
\end{equation}

\begin{figure}[ht]
 \centering
 \includegraphics [width=\linewidth]{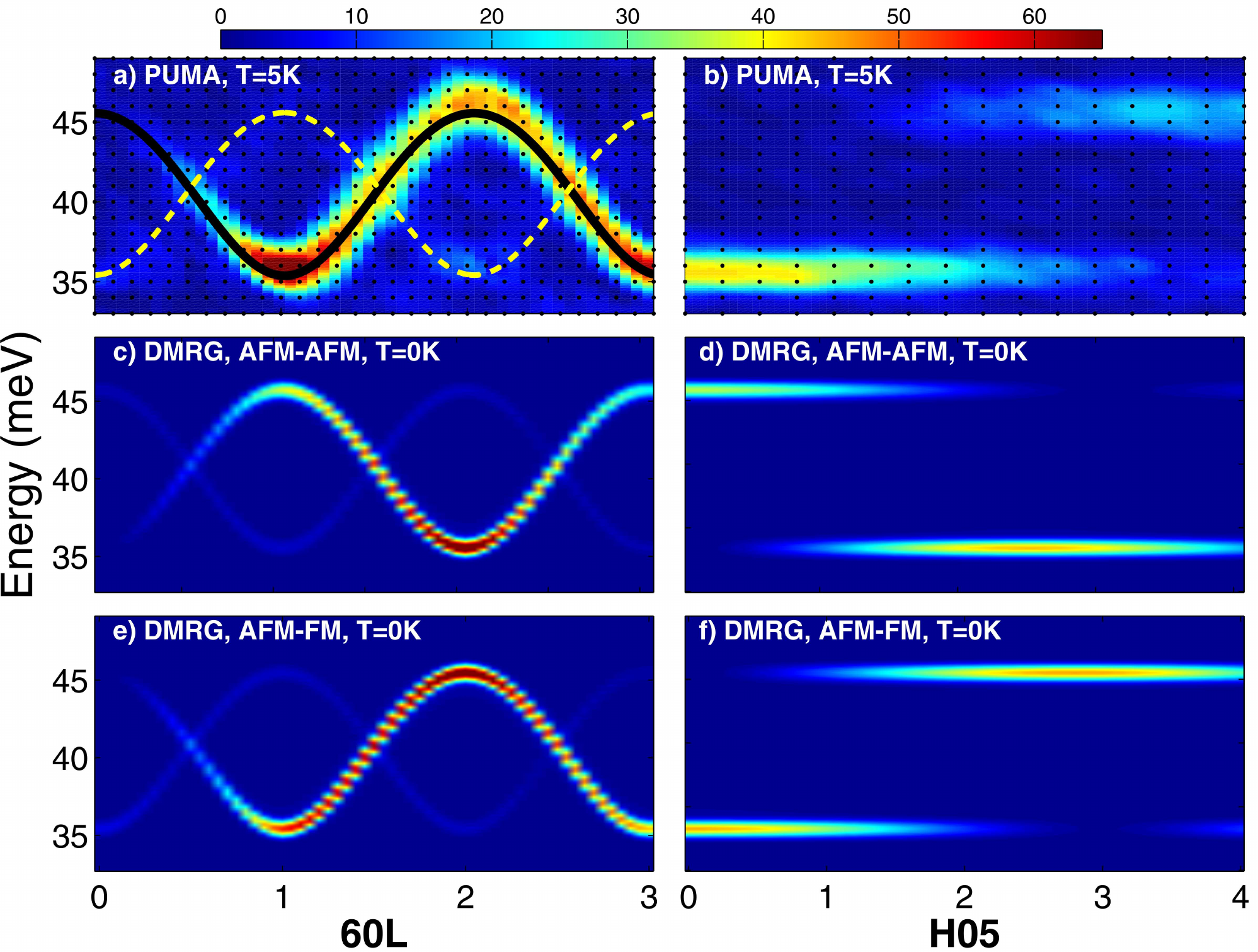}
 \caption{(Color online) Background-subtracted INS data along (a) (6,0,L) and (b) (H,0,5). The dashed and solid lines show the one-triplon dispersion to fifth order for the $J_1$-$J_2$ model with AFM-AFM ($J_{1} = 40.75$ meV, $J_{2} = 9.16$ meV) and AFM-FM ($J_1 = 40.92$ meV, $J_2=-11.97$ meV) interactions, respectively. DMRG results for the dynamic structure factor for the $J_1$-$J_2$ model with AFM-AFM interactions along (c) (6,0,L) and (d) (H,0,5), and AFM-FM interactions along (e) (6,0,L) and (f) (H,0,5). The anisotropic magnetic form factor of the Cu$^{2+}$ ions is taken into account \cite{Shamoto1993} and a resolution broadening is included.}
 \label{Fig2}
\end{figure}  

\par {\it{Deducing the Hamiltonian.$-$}}Figure\ \ref{Fig2}(a)-(b) presents INS data measured in the (H,0,L) plane at $T = 5$\,K. The magnetic excitation spectrum consists of two gapped branches which disperse along the L direction over the energy band $35.37 \pm 0.05$ meV to $45.56 \pm 0.05$ meV but are dispersionless along the H and K directions. The modes have the same periodicity and bandwidth, but are shifted with respect to each other by half a period and alternate in intensity. These results reveal that BaCu$_2$V$_2$O$_8$ is a highly dimerized 1D magnet where the dimers are coupled to form alternating chains along the $c$-axis while the coupling within the $ab$ plane is absent or negligibly small. The presence of a structure factor with two modes implies that these chains are not straight. Each mode is well reproduced by the one-triplon dispersion of an alternating chain \cite{a18,knett00a} assuming that either both interactions are AFM (dashed line in Fig.\ \ref{Fig2}(a)) or AFM and FM (solid line, Fig.\ \ref{Fig2}(a)). The extracted value of the alternating chain periodicity ($d$ = 4.04 $\pm$ 0.04 \AA) is the same for both modes and is half the $c$ lattice parameter. This periodicity corresponds to the alternating screw chain model ($J_1$-$J_2$), while the linear chain model ($J_{1}^{\ast}$-$J_2$) can be excluded because it would have a periodicity of $d$ = $c$ = $8.148$ \AA. Assuming that both the $J_{\mathrm{intra}}$ and $J_{\mathrm{inter}}$ interactions are AFM high-resolution energy scans at the dispersion minima and maxima were fitted using the fifth-order expansion of the alternating chain dispersion \cite{a18} and give the solution $J_{\mathrm{intra}}=40.75 \pm 0.02$ meV and $J_{\mathrm{inter}}=9.16 \pm 0.1$ meV. Equally good agreement was achieved for the AFM-FM model with exchange constants $J_{\mathrm{intra}}=40.92 \pm 0.01$ meV and $J_{\mathrm{inter}}=-11.97 \pm 0.1$ meV \cite{SM}. 

To distinguish between the alternating AFM-AFM and AFM-FM screw chain models, DMRG computations of the magnetic excitation spectra were performed. The results for the (6,0,L) and (H,0,5) directions at zero temperature are shown for the AFM-AFM (Fig.\ \ref{Fig2}(c)-(d)) and AFM-FM (Fig.\ \ref{Fig2}(e)-(f)) models. In both cases gapped modes are predicted, matching the experimental data in terms of energy and periodicity. However, only the AFM-FM model agrees with the observed intensity while the AFM-AFM chain is clearly wrong since the  intensities of the two modes are interchanged with respect to the experiment.  

Static magnetic susceptibility verifies this result. Figure\ \ref{Fig1}(b) shows the measured $\chi _{DC}$ for a magnetic field applied parallel and perpendicular to the $c$-axis. While these two directions have similar features, they have different amplitudes because of the anisotropic $g$-factor of Cu$^{2+}$ in this compound (caption of Fig.\ \ref{Fig1}). DMRG calculations of $\chi _{DC}$ were performed with the intrachain exchange constants fixed to the values obtained for the AFM-AFM and AFM-FM models. Best agreement is found for the AFM-FM model confirming the FM nature of the interdimer interaction. In addition the coupled dimer model \cite{a22,a23} was fitted to the data by varying the exchange constants and yields $J_{\mathrm{intra}} =39.8 \pm 0.13$~meV and $J_{\mathrm{inter}}=-9.87 \pm 2.64$~ meV again confirming the AFM-FM model. 

Our results reveal that BaCu$_2$V$_2$O$_8$ is an $S=1/2$ alternating screw chain with exchange paths $J_1$ and $J_2$  as predicted by band structure calculations \cite{a11,a16}. However, in contrast to these predictions and to all previous experimental work \cite{a12,a13,a14,a15} which assumed both interactions to be AFM, we demonstrate that the weaker interdimer coupling is FM. While we cannot determine which of the two exchange paths is FM, it is most likely that $J_1=J_{\mathrm{intra}}$ is AFM, while $J_2=J_{\mathrm{inter}}$ is FM. Indeed, band structure calculations predict that the super-superexchange path $J_1$ provides the strongest AFM interaction \cite{a11,a16} while the bridge angle of the $J_2$ Cu-O-Cu path is $94^{\circ}$ and is close to the crossover from AFM to FM according to the Goodenough-Kanamori-Anderson rules \cite{a20,Kanamori1959,Anderson1950}.

\par
\begin{figure}
\centering
\includegraphics [width=\linewidth]{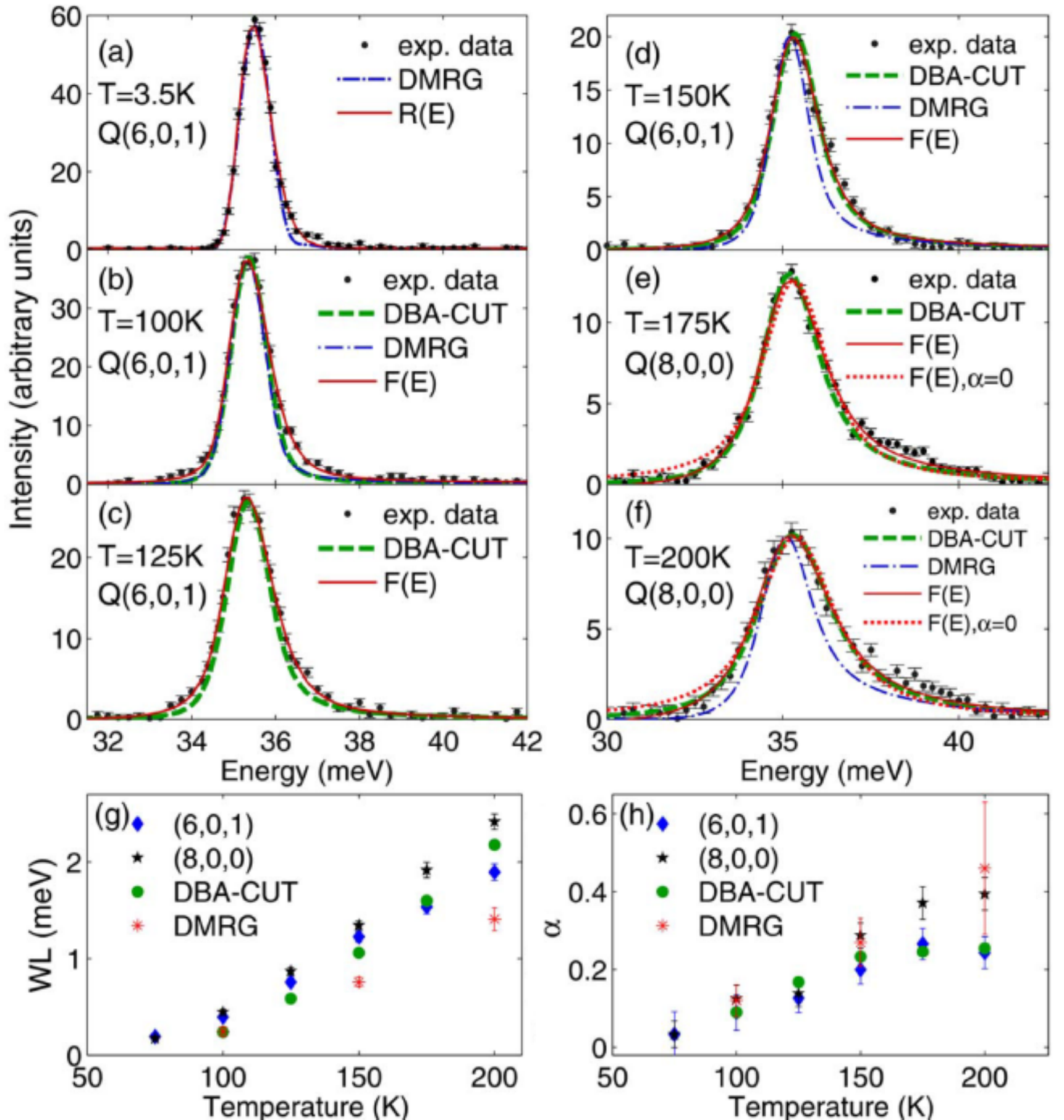}
\caption{(Color online) Background-subtracted constant wave-vector scans at (6,0,1) and (8,0,0) measured at (a) $T=3.5$ K (b) $T=100$ K (c) $T=125$ K (d) $T=150$ K (e) $T=175$ K (f) $T=200$ K. At $T=3.5$ K the excitations are resolution limited and the solid line gives the fit of the Gaussian function ($R(E)=G(W_G+\beta(E-E_0),E)$) where the Gaussian width $W_G$ has been replaced by $W_G \rightarrow W_G+\beta(E-E_0)$ which reproduces the asymmetric instrumental resolution function for finite $\beta$ \cite{SM}. At other temperatures the solid and dotted lines correspond to the fits of Eq.\ \eqref{eq: fitfunc} with $\alpha$ varied and $\alpha = 0$, respectively. The dashed and dash-dotted lines are results from DBA-CUT and DMRG for the AFM-FM model, respectively. Panels (g) and (h) show the temperature dependence of $W_L$ and $\alpha$ from fitting the experimental data and theoretical results.} 
\label{Fig3}
\end{figure}

 {\it{Strongly correlated behavior.$-$}} Now we turn to the question of whether BaCu$_2$V$_2$O$_8$ hosts strongly correlated behavior also at elevated temperatures. The alternating AFM-FM chain has received little experimental or theoretical attention since feasible physical realizations are rare. Thus, BaCu$_2$V$_2$O$_8$ provides the opportunity to investigate the effect of temperature on a new unexplored dimer system. This is achieved by performing energy scans at several temperatures up to 200 K (Fig.\ \ref{Fig3}) at the dispersion minima ((6,0,1) and (8,0,0)) where the deviations from symmetric Lorentzian behavior are most pronounced. Figure\ \ref{Fig3} shows that the excitations broaden with increasing temperature and at the highest temperatures the lineshape appears asymmetric and weighted towards the center of the band. By fitting the data at 175 K and 200 K to a symmetric Lorentzian $L(W_L,E)$ (where $W_L$ is the width and $E$ energy) convolved with the asymmetric instrumental resolution function \(R(E)\) given by the lineshape at base temperature (solid red line in Fig.\ \ref{Fig3}(a)), it is immediately clear that the lineshape of the excitations at these temperatures does not have the symmetric Lorentzian profile represented by the dotted red line in Figs. \ref{Fig3}(e)-(f). 

In order to capture the asymmetry, the conventional width $W_L$ in the Lorentzian function was replaced by ($W_L \rightarrow W_L + \alpha (E-E_0)$) where a finite $\alpha$ makes the lineshape asymmetric about the peak position $E_0$. Thus the new fitting function $F(E)$ is
\begin{equation}
F(E) ={A\cdot L\left(W_L+\alpha(E-E_0),E\right)\ast R(E)}.
\label{eq: fitfunc}
\end{equation}
Here $A$ denotes the peak intensity. The solid red lines in Figs.\ \ref{Fig3}(b)-(f) present our best fits of $F(E)$ to the experimental data and reveal that the lineshape of the excitations is asymmetric even down to 100 K. Figures\ \ref{Fig3}(g)-(h) display the extracted values of $W_L$ and of the asymmetry parameter $\alpha$ as a function of temperature and show that both increase with temperature. 

{\it{Comparison with theory.$-$}} To verify the experimentally observed asymmetric thermal lineshape broadening, we now compare it to theoretical results obtained by DMRG and DBA-CUT at finite temperatures for the AFM-FM model. Both approaches take account of the Gaussian resolution broadening but not of the asymmetry in the resolution function observed in the experiment. The DMRG results at $T=100$~K, 150 K, and 200 K reproduce the experimental data in Fig.\ \ref{Fig3}(b), (d), and (f) (dash-dotted blue line) assuming that the intradimer coupling changes slightly as temperature increases \cite{SM}. The dashed green line in Fig.\ \ref{Fig3}(b)-(f) represents the dynamic structure factors computed by DBA-CUT for 100~K, 125~K, 150~K, 175~K, and 200~K. Because the DBA-CUT peak positions are slightly offset from the experimental peaks at elevated temperatures \cite{SM} they were shifted for comparison to the experimental lineshapes. Both techniques clearly predict asymmetric lineshape broadening weighted towards higher energies at finite temperatures. These two very different theoretical approaches are in good quantitative agreement, with the DBA-CUT approach better able to resolve the lineshape, while the DMRG better obtains the peak position \cite{SM}.  When fitting the theoretical results using $F(E)$ and taking into account their resolution functions \cite{SM}, we extract the temperature dependence of $W_L$ and $\alpha$ as plotted in Figs.~\ref{Fig3}(g)-(h), showing good quantitative agreement with the experiment. This confirms the persistence of correlation effects in this system at elevated temperatures, and in addition shows that this effect is independent of the sign of the interdimer exchange coupling (see the SM \cite{SM} for a comparison of lineshapes of AFM-AFM and AFM-FM models). 

\par 
{\it{Summary.$-$}} Combining currently developed theoretical approaches and high-precision inelastic neutron scattering we quantitatively described the strongly correlated behavior at elevated temperatures in the 1D gapped dimer magnet BaCu$_2$V$_2$O$_8$ up to relatively high temperatures. Based on a customized fitting function the asymmetry could be reliably captured and parameterized. Our first key result is the very good agreement between the experimentally observed and the theoretically computed lineshapes obtained by the DMRG and the DBA-CUT approach which demonstrates accurate prediction of coherent behavior in quantum magnets. In this way, one can identify strongly correlated systems which retain their coherence at elevated temperatures. Our second key result is that we unambiguously established the relevant Hamiltonian of BaCu$_2$V$_2$O$_9$ revealing that it is a rare example of an alternating AFM-FM chain and correcting all previous results which assumed it to be an alternating AFM-AFM chain or an isolated dimer system \cite{a11,a12,a14,a15,a16}. This finding implies our third key result that strong correlations in dimerized quantum magnets at elevated temperatures are independent of the sign of the interdimer exchange coupling. Equipped with these techniques and insights, we anticipate future investigations to explore how strongly correlated behavior depends quantitatively on relevant parameters such as the dimension of the system, the size of the spins, and the statistics of the elementary excitations.

\begin{acknowledgments}
{\it Acknowledgments.} We would like to express our gratitude to our late colleague Prof. Thomas Pruschke for his unflagging support of this project and his incisive contributions as a physicist. This work is based upon experiments performed at the PUMA instrument operated by FRM II at the Heinz Maier-Leibnitz Zentrum (MLZ), Garching, Germany. We acknowledge the Helmholtz Gemeinschaft for funding via the Helmholtz Virtual Institute (Project No.\ VH-VI-521) and CRC/SFB 1073 (Project B03) of the Deutsche Forschungsgemeinschaft. We also thank D.A. Tennant and D. L. Quintero-Castro for helpful discussions.
\end{acknowledgments}

%%% Supplemental Material

\section{Supplemental Material}

\subsection{Values of interchain coupling, size of the gap and band width}
\par
\begin{figure}
 \centering
 \includegraphics [width=\linewidth]{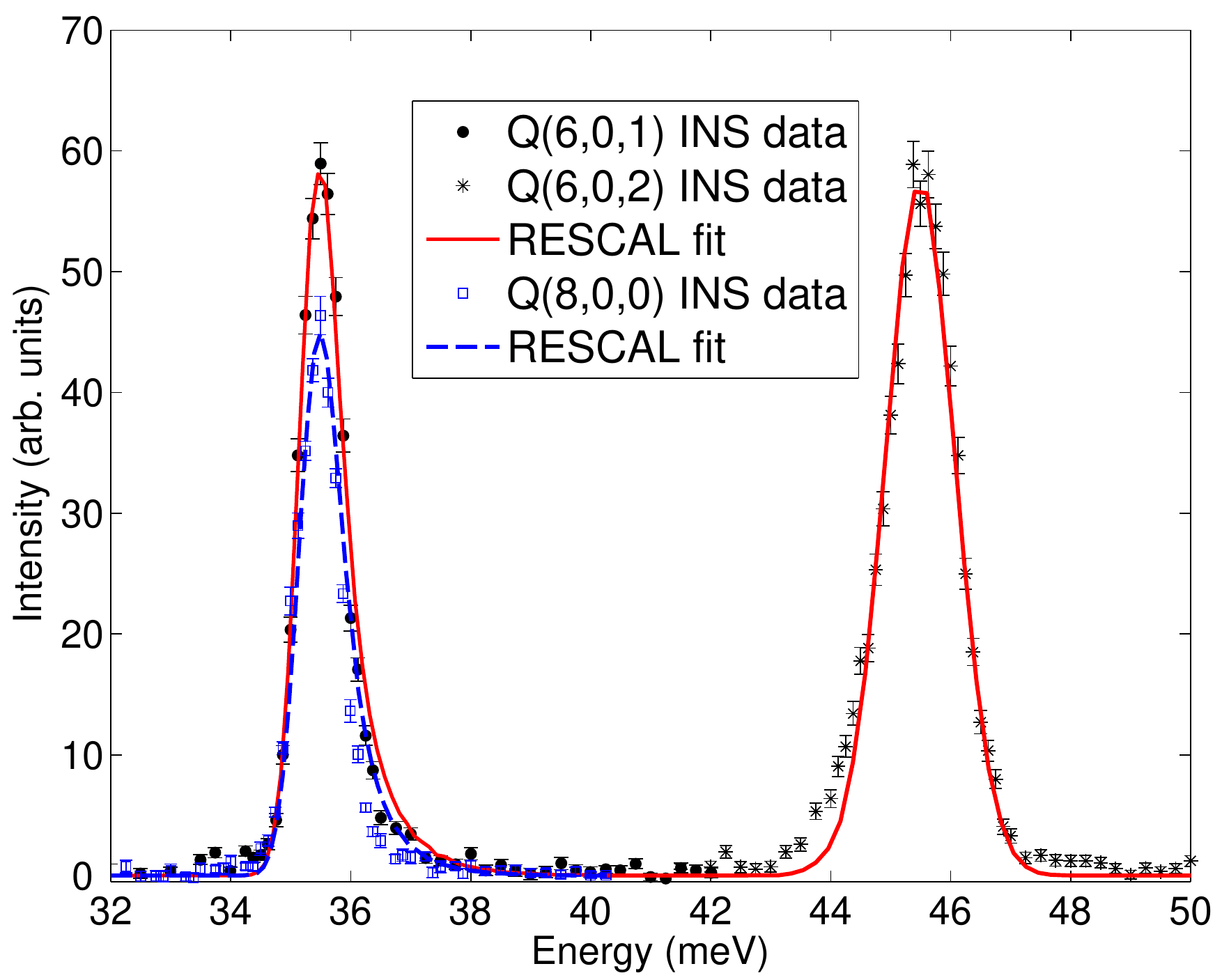}
 \caption{Energy scans at the dispersion minima (6,0,1), (8,0,0), and at the dispersion maximum (6,0,2). The solid red line represents the fit of the (6,0,1) and (6,0,2) data by using the fifth-order one-triplon dispersion relation convolved with the instrumental resolution function. The dashed blue line gives the fit of (8,0,0) using the same function.}

 \label{SMFig1}
\end{figure}

 In order to extract the values of the magnetic interchain coupling and to determine the size of the energy gap and the band width, the energy scans at the dispersion minima (6,0,1), (8,0,0), and the dispersion maximum (6,0,2) at the base temperature of $T = 3.5$~K were fitted by the one-triplon dispersion relation \cite{a18} convolved by the instrumental resolution function calculated by the RESCAL software 
\cite{RESCAL_misc}.
There are two modes in the experimentally observed magnetic excitation spectrum which have the same periodicity but are shifted with respect to each other by half a period. Because of this shift the two modes provide different solutions with opposite sign of the interdimer exchange coupling $J_2$. Figure\ \ref{SMFig1} shows the energy scans at (6,0,1) and (6,0,2) which are the dispersion minima and dispersion maxima of the intense mode along the (6,0,L) direction. The scans were measured at different fixed final wave vectors of $k_f =1.97$ \AA$^{-1}$ and $k_f =2.662$ \AA$^{-1}$ respectively giving different energy resolutions of 0.86~meV and 2~meV, as a result the peak at (6,0,2) is wider than the peak at (6,0,1). The solid red line gives the fit of one-triplon dispersion relation calculated to fifth order \cite{a18} convolved by the instrumental resolution using the RESCAL software and corresponds to the solution with $J_1=40.92$ meV and the ratio of $J_2/J_1= -0.2925$ implying that the interdimer coupling is $J_2=-11.97$~meV. The peaks are resolution limited with intrinsic widths of 0.01 meV while the observed asymmetric line shape of the peaks are caused by the instrumental resolution function in combination with the dispersion relation and are well reproduced by the RESCAL software. The instrumental resolution also affects the experimentally observed peak positions shifting the peak observed at the minimum and the maximum of the dispersion by $\approx$ 0.15 meV towards the center of the band. Indeed, the fifth-order one-triplon dispersion relation gives the peak positions of $E_{601}=35.37\pm0.05$~meV and $E_{602}=45.56\pm0.05$~meV which are in agreement with the experimentally observed values of $E_{601}=35.5\pm0.05$~meV and $E_{602}=45.49\pm0.05$~meV within the error of $\approx$ 0.15 meV due to the resolution effects. 
\par The dashed blue line in Fig.\ \ref{SMFig1} shows the fit of the dispersion minimum at (8,0,0) by the second mode which corresponds to the values of $J_1=40.74$ meV and $J_2/J_1= 0.227$ implying  that the interdimer coupling is antiferromagnetic. A more accurate solution was extracted by fitting the energy scans at (6,0,1) and (6,0,2) by the fifth-order one-triplon dispersion shifted by half a period. This yielded $J_1=40.75$ meV and $J_2/J_1= 0.225$.    
            
\subsection{Analytical description of the instrumental resolution function}
\par
\begin{figure}
 \centering
 \includegraphics [width=\linewidth]{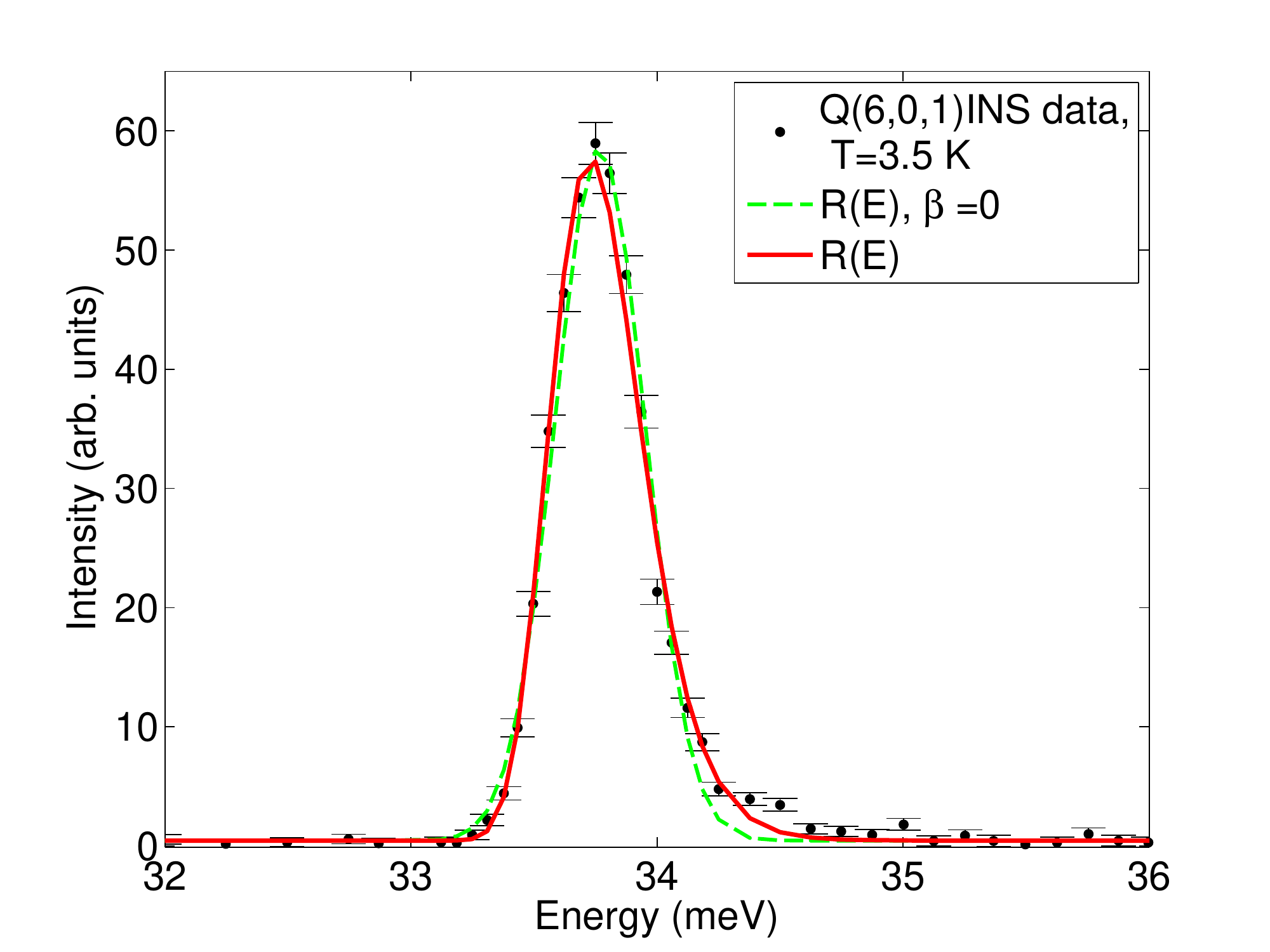}
 \caption{Energy scans at the dispersion minimum (6,0,1). The solid red line represents the fit at (6,0,1) by using the function $R(E)$ with parameters $W_G=0.89\pm0.03$~meV, $\beta=0.1 \pm 0.04$.}
 \label{SMFig2}
\end{figure}

 As found in the previous section numerical computations of the instrumental resolution function using the RESCAL software show that the data at base temperature are described entirely by the resolution function. Conventionally the Gaussian function is used to characterize the instrumental resolution function. The energy scans at dispersion minima at (8,0,0) and (6,0,1) at $T=3.5$~K were fitted by using the conventional Gaussian function (dashed green line in Fig.\ \ref{SMFig2}). The extracted full width at half maximum ($W_G$) was found to be of $W_G=0.863\pm0.010$ meV for the instrument settings that were used and is associated with the instrumental resolution broadening. This value is in good agreement with the calculated value of 0.74~meV and the value of $W_G=0.86$~meV that was used in the DMRG and DBA-CUT calculations at finite temperatures to take into account the resolution broadening. 
\par However the conventional Gaussian function does not reproduce the asymmetry of the line shape of the instrumental resolution function which was experimentally observed to be weighted towards the center of the band and was numerically reproduced using the RESCAL software (see previous section). At finite temperatures, where the intrinsic line shape broadens and can also become asymmetric, contributions from both the resolution and this intrinsic line shape broadening combine to produce the observed asymmetric shape of the peak. Therefore, it is important to distinguish the asymmetry due to the resolution from that due to the intrinsic line shape. This was achieved by introducing an asymmetric Gaussian function to analytically describe the instrumental resolution function. This function is based on a normalized Gaussian, where $W_G$ is replaced by \(W_G+\beta(E-E_0)\) and $\beta$ describes the asymmetry of the instrumental resolution function  
\begin{multline}
R(E)={Gaussian\left(W_G+\beta(E-E_0),E\right)} \\ 
={\frac{ \exp
\left(
\frac{-(E-E_0)^2}
{2\biggl(
\frac{W_G}{2\sqrt{2 \ln(2)}}+\beta(E-E_0)
\biggr)^2}
\right)}
{\sqrt{2\pi}
\biggl(
\frac{W_G}{2\sqrt{2 \ln(2)}}+\beta(E-E_0)
\biggr)}}.
\label{eq: fitfunc}
\end{multline}
Here $E_0$ denotes the peak position.

\par To quantitatively describe the asymmetry of the instrumental resolution function, the energy scans at the dispersion minima (6,0,1) and (8,0,0) at the base temperature of 3.5~K were fitted by the function $R(E)$ (solid red line in Fig.\ \ref{SMFig2}) and the extracted values of $W_G=0.89 \pm 0.03$~meV and $\beta =0.1 \pm 0.04$ were fixed in the analysis at finite temperatures. 

 \subsection{Description of the fitting function $F(E)$}
  \par At finite temperatures the intrinsic line shape of the excitations broadens and becomes asymmetric. An asymmetric Lorentzian function is used to describe it where the Lorentzian width $W_L$ is replaced by \(W_L+\alpha(E-E_0)\) and $\alpha$ parameterizes the asymmetry. This function has to be convolved with the resolution function $R(E)$ determined at base temperature to reproduce the observed line shape

\begin{multline}  
F(E) =\frac{A}{\pi}\cdot L\left(W_L+\alpha(E-E_0),E\right)\ast R(E) \\ 
=\frac{A}{\pi}\cdot\int\limits_{-\infty}^{\infty}\mathrm{d}t\frac{\frac{W_L\left(T\right)+\alpha\left(T\right)\left(t-E_0\left(T\right)\right)}{2}}
{\left(t-E_0\left(T\right)\right)^2+\left(\frac{W_L\left(T\right)+\alpha\left(T\right)\left(t-E_0\left(T\right)\right)}{2}\right)^{2}}  
\\
\times \frac{\exp\left(\frac{-(E-t)^2}{2\biggl(\frac{W_G}{2\sqrt{2\ln(2)}}+\beta(E-t)\biggr)^2}\right)}{\biggl(\frac{W_G}{2\sqrt{2\ln(2)}}+\beta\left(E-t\right)\biggr)} 
\label{eq: fitfunc_convolution}
\end{multline}
The function $F(E)$ was implemented using the MatLab 2012b Software and the infinite  integral was replaced by the definite integral over the interval [$-150$ meV, 150~meV].

\par The energy scans at the dispersion minima (6,0,1) and (8,0,0) at finite temperatures were fitted using the function $F(E)$ with the parameters $W_G$ and $\beta$ of the resolution function $R(E)$ fixed to the the corresponding values extracted at base temperature. The parameters of $W_L$ and $\alpha$ were varied to describe the intrinsic asymmetric thermal line shape broadening of the magnetic excitations.

\subsection{DMRG calculations}

The excitation spectra were computed directly in the frequency domain for one-dimensional systems of finite size and open boundary conditions. To this end, we employed DMRG-based Chebyshev expansions at zero~\cite{Weisse2006,Holzner2011, Braun2014, Wolf2015} and finite temperature \cite{Tiegel2014, Tiegel2015}. At $T=0$ K, the quantity of interest is the longitudinal dynamical structure factor
\begin{subequations}
\begin{align}
   & S_{zz}^{T=0}  (\omega, Q) = \sum_n \left| \langle n | S^z_Q | 0 \rangle \right|^2 \delta(\omega - (E_n-E_0)) \\
   & = \langle 0 | \,S^z_{-Q} \, \delta(\omega - (H-E_0)) \, S^z_Q \,| 0  \rangle \label{eq: spec_func} \\
   &=       \sum_{j,\,l}  \text{e}^{2\pi i\, Q \cdot (R_l-R_j)}    \underbrace{\langle 0 | \,S^z_l \, \delta(\omega - (H-E_0)) \, S^z_j \,| 0  \rangle}_{=: G_{l,j}(\omega)}.  \label{eq: spec_func_spatial}
\end{align}
\end{subequations}
Here $| n \rangle$ and $E_n$ denote the eigenstates and eigenvalues of the Hamiltonian $H$. For the Fourier transform of the spin operator
\begin{align}
  S^z_{Q} = \frac{1}{\sqrt{N_{\rm atom}}} \sum_j \text{e}^{-2\pi i\, Q \cdot R_j} \, S^z_j \label{eq: Fourier}
\end{align}
we use the real positions $R_j$ of $N_{\rm atom}=80$ copper atoms in BaCu$_2$V$_2$O$_8$. It is important to take account of the crystal structure since we could demonstrate that the weaker dispersion in Fig.\ 2(a) of the main text is engendered by the screw-chain geometry of the compound. The momentum $Q$ is specified by Miller indices $Q= \rm (H,K,L)$ and $S^z_j$ is the $z$ component of the local spin operator acting at site~$j$. We perform an MPS-based expansion of the dynamical spin structure factor in Chebyshev polynomials~\cite{Weisse2006,Holzner2011, Braun2014, Wolf2015}. Note that by subsequently convolving the Chebyshev expansion with the Jackson kernel \cite{Weisse2006} we introduce a nearly Gaussian broadening corresponding to the experimental resolution. This also helps to avoid Gibbs oscillations occurring as a consequence of the finite expansion order. Higher expansion orders enhance the resolution since the broadening is inversely proportional to the expansion order. For further details of this approach refer to Ref.\,\cite{Tiegel2015} and the references therein. 

Equations~\eqref{eq: spec_func} and~\eqref{eq: spec_func_spatial} now offer two different schemes for the computation of the spectral function. In the first case of Eq.~\eqref{eq: spec_func}, $Q$ is specified prior to one single calculation in momentum space. More flexibility is provided by the scheme suggested in Eq.~\eqref{eq: spec_func_spatial}. Here the dynamical correlation functions $G_{l,j}(\omega)$ are computed individually in real space giving access to arbitrary momenta in the postprocessing stage, at the expense of an increasing computational effort by a factor $\sim N_{\rm atom}$. Moreover, we exploit the reflection symmetry of the system for $G_{l,j}(\omega)$. The latter computation scheme is therefore advantageous in order to obtain the DMRG results along various directions in $Q$ space shown in Fig.\,2 of the main text. For these zero-temperature data we retain a maximal internal MPS bond dimension of $m=150$. In each Chebyshev iteration the error resulting from the variational compression is $\epsilon_{\rm compr} < 10^{-9}$. Since there are two screw chains with different winding orientation in each BaCu$_2$V$_2$O$_8$ unit cell, the results are obtained as a superposition of both screw chains. 

In the $T>0$ case, we exploit a Liouville-space formulation for the frequency-space dynamics which can be implemented as a Chebyshev expansion for the Liouville operator in a DMRG framework \cite{Tiegel2014,Tiegel2015}. This approach is used to approximate the finite-temperature dynamical spin structure factor given by
\begin{align*}
S_{zz}^{T>0} (\omega, Q) &= \frac{1}{Z} \, \sum_{n,m} \, \text{e}^{-E_n/(k_\mathrm{B}\,T)} \langle m | S^z_{-Q} | n \rangle \times \\
  &\quad \quad \times \langle  n | S^z_Q | m \rangle \,\delta\left( \omega - (E_m-E_n)\right). 
\end{align*}
Here $Z=\sum_n \text{e}^{-E_n/(k_\mathrm{B}\,T)}$ denotes the canonical partition function and $k_\mathrm{B}$ the Boltzmann constant. At $T>0$, it is computationally more expensive to compute the expansion coefficients that are also called Chebyshev moments. For details on this issue refer to Ref.~\cite{Tiegel2015}. We therefore calculate only about 1000 moments using a higher MPS bond dimension ($m=250$) than at $T=0$, leading to a compression error of $\epsilon_{\rm compr} \lesssim 10^{-4}$. The computation proceeds directly in momentum space, i.e., by means of the finite-temperature analogue of Eq.~\eqref{eq: spec_func}. Subsequently, we extrapolate these computed moments with linear prediction~\cite{Makhoul1975} in order to obtain a higher resolution allowing for a direct comparison to the experiments in Fig.\,3 of the main text. For time-dependent DMRG, linear prediction is an established method \cite{White2008, Barthel2009} and it has recently also been extended to determine Chebyshev moments \cite{Ganahl2014, Wolf2014a}. In order to assess the quality of the extrapolation at a given temperature, we varied the input parameters for the linear prediction, e.g., the number of computed Chebyshev moments and the training interval. These data sets produced by linear prediction were then fitted by the function in Eq.~\eqref{eq: fitfunc_convolution}. The resulting fit parameters for the Lorentzian width $W_L$ and asymmetry $\alpha$ also displayed a slight dependence on the fit interval at a fixed temperature. The error estimates shown in Fig.\,3(g)-(h) of the main text represent the maximal deviation found in our analysis.  
At $T>0$, we only use the atom positions of a single screw chain consisting of $N_{\rm atom}=40$ Cu$^{2+}$ ions for the Fourier transform in Eq.~\eqref{eq: Fourier} since the effect of a different winding orientation is negligible.

\subsection{Diagrammatic Br\"uckner approach}

\begin{figure}[]
\centering
\vspace*{12pt}
\includegraphics[width=0.6\columnwidth]{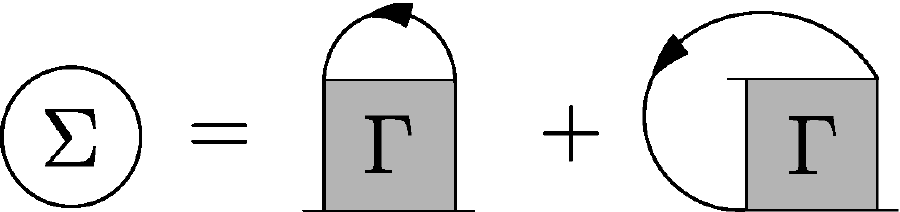}
\caption{Self-energy diagrams in leading order in $\exp \left(-\Delta/(k_\mathrm{B}T)\right)$. The first diagram generates Hartree-like diagrams and the second the Fock-like diagrams with a renormalized interaction given by
the scattering amplitude $\Gamma$.}
\label{fig.self_energy}
\end{figure}

\begin{figure}[]
\centering
\includegraphics[width=0.99\columnwidth]{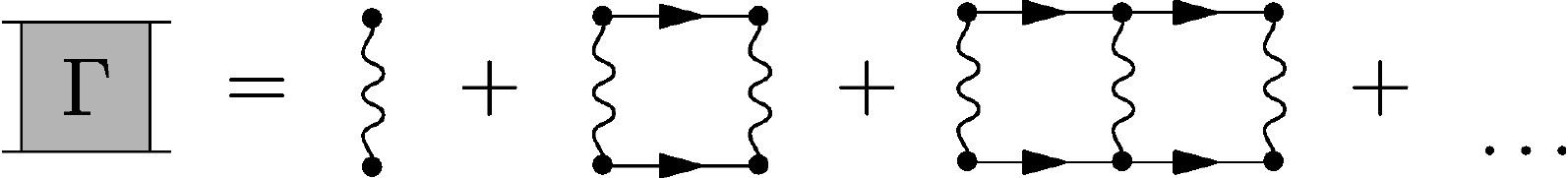}
\caption{Definition of the scattering amplitude $\Gamma$.}
\label{fig.ladder2}
\end{figure}

The diagrammatic Brückner approach was first introduced for spin systems at $T=0$ in Ref.~\cite{kotov98a} and extended to finite temperature fluctuations in combination with effective models in Refs.\ \cite{fause14,fause15}. The systematic control parameter of the approach is the low density of thermally excited hardcore bosons
which is proportional to $\exp \left(-\Delta/(k_\mathrm{B}T)\right)$ where $\Delta$ is the energy gap.\\
The approach applies in any dimension and works directly in frequency and momentum space. The continuation from Matsubara frequencies to real frequencies is performed analytically.\\
Technically, we start from an effective model, which conserves the number of quasi-particles in the system.
This effective model is computed by a continuous unitary transformation (CUT) \cite{wegne94, glaze93, knett00a, knett03a,kehre06, fisch10a, krull12}. The effective Hamiltonian is written in terms of triplon operators acting on the ground state of singlets on the dimers \cite{knett03a}. \\
In first order of the parameter $x=J_\mathrm{inter}/J_\mathrm{intra}$, the effective Hamiltonian in terms of triplon operators is given by
\begin{subequations}
\begin{align}
H_\mathrm{eff} &= E_0 + \sum\limits_i \sum\limits_\alpha t_{i,\alpha}^\dagger t_{i,\alpha}^{\phantom\dagger}  \\
  &- \frac{x}{4} \sum\limits_i \sum\limits_\alpha t_{i,\alpha}^\dagger t_{i+1,\alpha}^{\phantom\dagger}  + \mathrm{h.c.}\\
  &+ \frac{x}{4} \sum\limits_i \sum\limits_{\alpha \neq \phi} t_{i,\alpha}^\dagger t_{i,\phi}^{\phantom\dagger} t_{i+1,\phi}^\dagger t_{i+1,\alpha}^{\phantom\dagger} - t_{i,\alpha}^\dagger t_{i,\phi}^{\phantom\dagger} t_{i+1,\alpha}^\dagger t_{i+1,\phi}^{\phantom\dagger}  \label{eq: interaction_in_effective_Hamiltonian} \\
  &+ \mathcal{O}(x^2).
\end{align}
\end{subequations}
Here $\alpha, \phi \in \lbrace x,y,z\rbrace$ denote the possible flavors of the excited triplons and $E_0$ is the ground state energy. The operators $t_{i,\alpha}^{(\dagger)}$ create/annihilate a triplon of flavor $\alpha$ on site $i$. They are hardcore bosons because on a single dimer only one excitation is allowed at maximum.
Note that the index $i$ labels dimers.\\ Since $x \approx 0.3$ in the case of BaCu$_2$V$_2$O$_8$, an order $6$ calculation for the CUT is sufficient to capture all quantum fluctuations at zero temperature quantitatively, i.e., $H_\mathrm{eff}$ is known up to all terms in order $x^6$. \\
On top of this effective Hamiltonian, we apply diagrammatic perturbation theory to account for the thermal effects in the spectrum.
The idea is to treat the hardcore bosons as normal bosons, but with an infinite on-site interaction. \\
The quantity of interest is the dynamic structure factor which is related to the imaginary part of the Green function by means of the fluctuation-dissipation theorem
\begin{align}
\label{eq.fluctuation_dissipation}
S^{T>0}_{zz}(p, \omega) = \frac{1}{1-e^{-\omega/(k_\mathrm{B} T)}} \frac{1}{\pi} \mathrm{Im} \left[ G^{zz}(p, \omega) + G^{zz}(p, -\omega) \right].
\end{align}
The imaginary part of the Green function is calculated by the Br\"uckner approach by means of the single particle self-energy. In terms of diagrams, we have to sum all contributions displayed in Fig.\ \ref{fig.self_energy}. 
Formally this translates to the expression
\begin{align}
\label{eq.self_energy1}
\Sigma^{\alpha \alpha}(P) = \frac{1}{N} \sum\limits_\phi \sum\limits_{K} (1+\delta_{\alpha,\phi}) G_0^{\phi \phi}(K) \Gamma^{\alpha,\phi}(P+K),
\end{align}
where $N$ denotes the total number of sites, $P$ and $K$ are 2-momenta, i.e.,\ $P = \left( p, i \omega_p \right)$, and $G^{\phi \phi}_0$ are the bare Green functions.
The scattering amplitude $\Gamma$ is the effective interaction between the hardcore bosons at finite temperature. Its graphical representation is given in Fig.\ \ref{fig.ladder2}. The interaction vertices in Fig.\ \ref{fig.ladder2} represent the local repulsion $U$, which is sent to infinity to realize the hardcore property. The scattering amplitude $\Gamma$ can be calculated by using the Bethe-Salpeter equation,
\begin{align}
\label{eq.bethe_salpeter}
\Gamma^{\alpha,\phi}(P) = \lim\limits_{U \rightarrow \infty} \frac{\frac{U k_\mathrm{B} T}{N}}
{1 + \frac{U k_\mathrm{B} T}{N} \sum\limits_K G_0^{\alpha \alpha}(P+K) G_0^{\phi \phi}(-K)}.
\end{align}
To also take the additional interactions in Eq.\ \eqref{eq: interaction_in_effective_Hamiltonian} into account, we apply a self-consistent Hartree-Fock decoupling similar to Ref.\ \cite{strei15}. Hence we decouple all quartic interactions other than the infinite local repulsion in the effective Hamiltonian according to
\begin{equation}
\begin{aligned}
t_{i,\alpha}^\dagger t_{i+d_1,\phi}^\dagger  t_{i+d_2,\gamma}^{\phantom\dagger} t_{i+d_3,\xi}^{\phantom\dagger}
&\approx
\left\langle t_{i,\alpha}^\dagger t_{i+d_2,\gamma}^{\phantom\dagger} \right\rangle t_{i+d_1,\phi}^\dagger t_{i+d_3,\xi}^{\phantom\dagger} \\
&+ \left\langle t_{i,\alpha}^\dagger t_{i+d_3,\xi}^{\phantom\dagger}\right\rangle  t_{i+d_1,\phi}^\dagger t_{i+d_2,\gamma}^{\phantom\dagger}  \\ 
&+ \left\langle t_{i+d_1,\phi}^\dagger t_{i+d_2,\gamma}^{\phantom\dagger} \right\rangle t_{i,\alpha}^\dagger t_{i+d_3,\xi}^{\phantom\dagger}   \\
&+ \left\langle t_{i+d_1,\phi}^\dagger t_{i+d_3,\xi}^{\phantom\dagger}\right\rangle t_{i,\alpha}^\dagger t_{i+d_2,\gamma}^{\phantom\dagger} \\
&+ \mathrm{const}.
\end{aligned}
\end{equation}
The decoupling has no effect on the imaginary part of the self energy but it shifts the peak positions slightly. A more sophisticated approach, which is subject of ongoing research, is to include the additional interaction in the Bethe-Salpeter equation, which yields the exact contribution of the additional interaction in $\exp \left(-\Delta/(k_\mathrm{B}T)\right)$.\\
Technically, all formulas are implemented on a finite size grid in frequency and momentum space. Convolutions are performed using fast Fourier algorithms of the FFTW library \cite{FFTW05}. Furthermore, the Green function is calculated self-consistently by replacing the bare Green function $G_0$ in Eq.\ \eqref{eq.self_energy1} and \eqref{eq.bethe_salpeter} by the interacting Green function $G$.

\subsection{Comparison of the theoretical results}
\begin{figure}[]
\centering
\includegraphics[width=0.99\columnwidth]{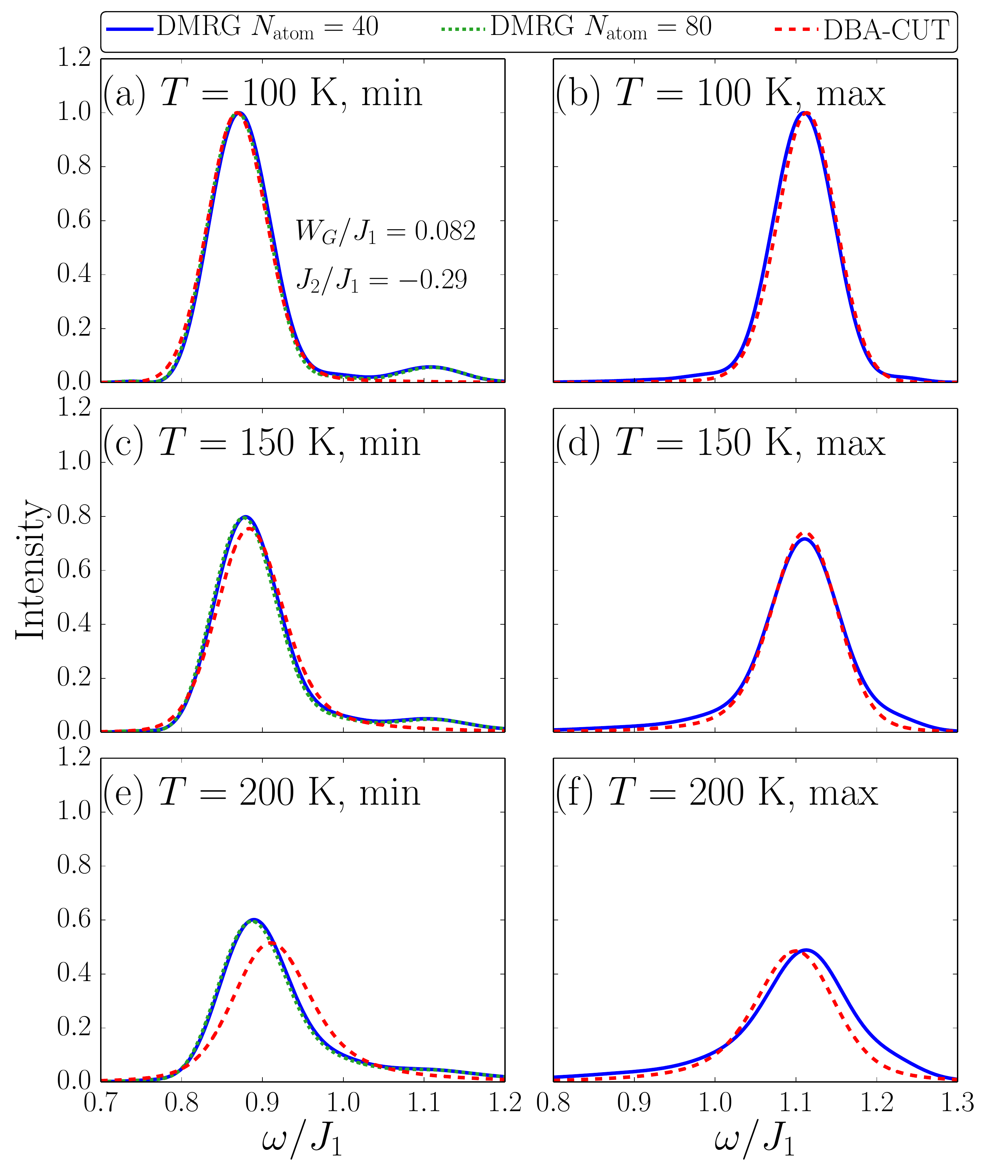}
\caption{Comparison of finite-temperature results obtained by the diagrammatic Br\"uckner approach and DMRG at the minimum (left column) and maximum (right column) of the single-triplon dispersion ($J_1=40.92$ meV).  }
\label{fig: comparison}
\end{figure}

Finally, we compare the finite-temperature results obtained by the diagrammatic Br\"uckner approach to the DMRG calculations in Fig.\ \ref{fig: comparison}, which shows the theoretical data for a lower resolution than in Fig.\ 3 of the main text. At the resolution we are comparing the results to each other, the DMRG-based computation of the Chebyshev moments is well controlled, so that we do not need to perform a linear prediction in the Chebyshev moments here. At the lowest temperature, $T=100$ K, in Fig.\,\ref{fig: comparison}(a)-(b), the peak heights are fixed to one. These scaling parameters are also kept at higher temperatures allowing for a direct comparison of the theoretical approaches. Concerning the evolution of the line shape with temperature, there is very good agreement of the two approaches up to $T=150$~K.  At $T=200$ K, there is a slight deviation in the peak positions, which can be explained by the low-temperature approximation inherent to the diagrammatic Br\"uckner approach. The leading order $\exp \left(-\Delta/(k_\mathrm{B}T)\right)$ is captured exactly but the shift is an effect $\propto \exp \left(-2\Delta/(k_\mathrm{B}T)\right)$ so that deviations occur \cite{fause14}.\\ In the left column of Fig.\ \ref{fig: comparison}, we also show DMRG results for different system sizes $N_{\rm atom}=40$ and~80. Since these two DMRG curves are very close to each other, finite-size effects seem to be negligibly small. Moreover, the line shape is not significantly altered by adopting a one-dimensional Fourier transform, which does not take account of the real atom positions, as used in the Br\"uckner approach.\\
We conclude that at the resolution shown in Fig.\ \ref{fig: comparison} both theories show an excellent agreement for the shape, width, and temperature-dependence of height and good agreement on the position at low temperatures. Deviations between both approaches only occur for very high resolutions as required for the quantitative analysis of the experiment. This leads to the slightly different results for width and asymmetry displayed in Fig. 3 (g) and (h) in the main text.

 \subsection{Temperature dependence of the intradimer magnetic exchange coupling $J_1$}
\begin{figure*}
 \centering
 \includegraphics [width=0.94\linewidth]{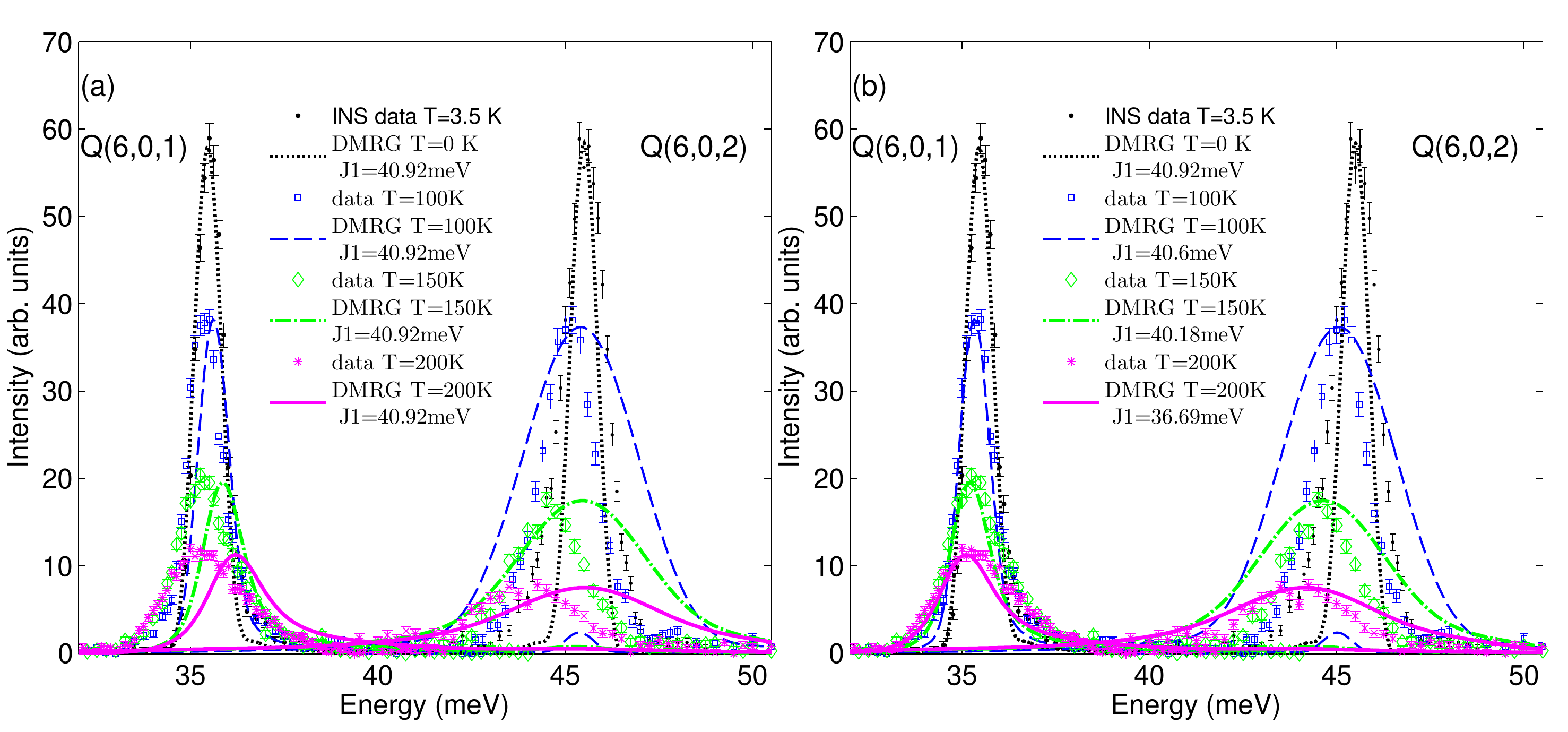}
 \caption{Energy scans at the dispersion minimum (6,0,1) and dispersion maximum (6,0,2) at temperatures of $T=3.5$~K, 100~K, 150~K, and 200~K compared to the corresponding DMRG calculations: (a) The DMRG calculations were scaled by using the value of $J_1 = 40.92$~meV at all temperatures. (b) The DMRG calculations were scaled by using the temperature dependent values of $J_1 = 40.92$~meV, 40.6~meV, 40.18~meV, 39.69~meV at $T=0$~K, 100~K, 150~K, and 200~K, respectively.}
 \label{SMFig6}
\end{figure*}

\par Since BaCu$_2$V$_2$O$_8$ is measured over a large temperature range, it is possible that small changes in the values of the exchange interactions $J_1$ and $J_2$ occur as the lattice distorts with increasing temperature. As a consequence, at finite temperatures the peak positions are affected by the combination of two factors: i) the energy shift due to thermal effects on the magnetic system ii) the energy shift due to changes of the magnetic exchange interactions $J_1$ and $J_2$. The DMRG takes into account the thermal effects and predicts a corresponding energy shift. However the DMRG does not predict the temperature dependence of the magnetic exchange interactions as the ratio of $J_2/J_1=-0.29$ was assumed to be temperature independent (see Fig.\ \ref{fig: comparison}). For the calculations, the intradimer coupling was set to unity and thus the energies were obtained in units of $J_1$.

\par Figure \ref{SMFig6}(a) shows the DMRG calculations, which are scaled to meV units using only the base temperature value of $J_1 =40.92$~meV (assuming temperature-independent interactions), in comparison to the experimental data. 
Note that the peak positions from the DMRG calculations can already be determined reliably at a lower resolution than the experimental one. Therefore, the DMRG data at (6,0,2) are not as highly resolved as at (6,0,1).
The DMRG calculations display an offset with respect to the experimental data that increases with temperature. The simulated DMRG data at both the dispersion minima and dispersion maxima are found at higher energy with respect to the experimental data. Therefore this shift cannot be attributed to thermal effects.
\par Using the DMRG calculations, the predicted positions of the center of the band (CB) for the temperatures $T=100$~K, 150~K, and 200~K in units of $J_1$  were compared to the experimentally observed center of the band at the corresponding temperatures. This information was used to extract the values of the magnetic exchange interactions $J_1$ as a function of temperature which are listed in Table\,\ref{tab: Jintra}.  Figure \ref{SMFig6}(b) presents the DMRG calculations which were scaled to meV units using the deduced values of $J_1(T)$ plotted over the experimental data. Using this scaling, the DMRG calculations at (6,0,1) and (6,0,2) are in a good agreement with the experimental data and the remaining tiny shift of the simulated peaks with respect to the experimental peak position positions (by$\approx$0.15 meV) is caused by resolution effects (see first section of this Supplemental Material). 

\begin{table}
\begin{ruledtabular}
\begin{tabular}{|c|c|c|c|}
		Temperature &CB experiment & CB DMRG \linebreak & $J_1(T)$ \\
		(K) & (meV) & ($E/J_1$) & (meV) \\
		\hline
    3.5 &40.495 $\pm$ 0.03&0.98961&40.92 $\pm$ 0.03\\
		\hline
    100 &40.19 $\pm$ 0.05&0.99&40.6 $\pm$ 0.05\\
		\hline
    150 &39.92 $\pm$ 0.05&0.9935&40.18 $\pm$ 0.05\\
		\hline
    200 &39.65 $\pm$ 0.1&0.999&39.69 $\pm$ 0.1 \\
\end{tabular}
\caption{Temperature-dependent values for the center of the band (CB) determined from the experiment and DMRG used to determine $J_1(T)$.}
\label{tab: Jintra}
\end{ruledtabular}
\end{table} 
     
\begin{figure}
 \centering
 \includegraphics [width=0.95\linewidth]{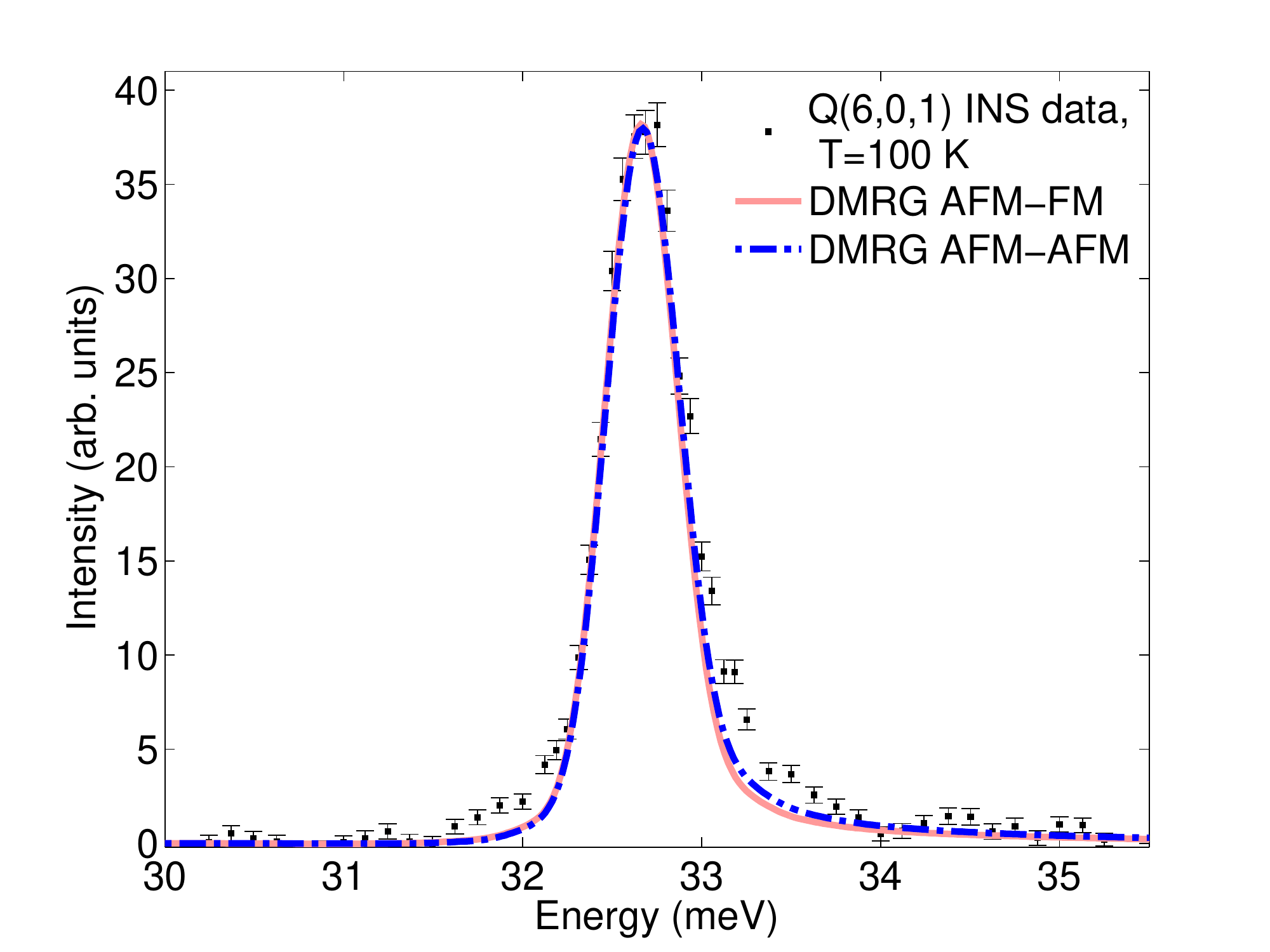}
 \caption{Energy scan at the dispersion minimum (6,0,1) compared to DMRG calculations for the AFM-FM model (solid line) and for the AFM-AFM model (dashed-dotted line) at $T=100$~K.}
\label{SMFig7}
\end{figure}     
     
 \subsection{Comparison of the AFM-AFM and AFM-FM models at finite temperatures}
Next, we compare the effect of the FM and AFM interdimer exchange coupling $J_2$ on the asymmetric line shape broadening.  The spectral function at the dispersion minimum was computed by DMRG for both the AFM-AFM and AFM-FM model at $T=100$~K. The results are plotted over the corresponding experimental data at (6,0,1) in Fig.\ \ref{SMFig7} and reveal that both models predict an almost identical asymmetric thermal line shape broadening at $T=100$~K.     

%\clearpage

%

\end{document}